\magnification 1200
\centerline {\bf  On the Mathematical Structure of  Quantum Measurement Theory}
\vskip 1cm
\centerline {{\bf by Geoffrey Sewell}\footnote*{e-mail: g.l.sewell@qmul.ac.uk}}
\vskip 0.5cm
\centerline {\bf Department of Physics, Queen Mary, University of London}
\vskip 0.5cm
\centerline {\bf Mile End Road, London E1 4NS, UK}
\vskip 1cm
\centerline {\bf Abstract}
\vskip 0.5cm
We show that the key problems of quantum measurement theory, namely the reduction of 
the wave-packet of a microsystem and the specification of its quantum state by a 
macroscopic measuring instrument, may be rigorously resolved within the traditional 
framework of the quantum mechanics of finite conservative systems. The argument is 
centred on the generic model of a microsystem, $S$, coupled to a finite macroscopic 
measuring instrument ${\cal I}$, which itself is an $N$-particle quantum system. The 
pointer positions of ${\cal I}$ correspond to the macrostates of this instrument, as 
represented by orthogonal subspaces of the Hilbert space of its pure states. These 
subspaces, or \lq phase cells\rq , are the simultaneous eigenspaces of a set of coarse 
grained intercommuting macro-observables, $M$, and, crucially, are of astronomically 
large dimensionalities, which increase exponentially with $N$. We formulate conditions 
on the conservative dynamics of the composite $(S+{\cal I})$ under which it yields both 
a reduction of the wave packet describing the state of $S$ and a one-to-one 
correspondence, following a measurement, between the observed  pointer position of 
${\cal I}$ and the resultant eigenstate of $S$; and we show that these conditions are 
fulfilled by the finite version of the Coleman-Hepp model.
 \vfill\eject
\centerline {\bf 1. Introductory Discussion.} 
\vskip 0.3cm
The quantum theory of measurement is concerned with the determination of the state of a 
microsystem, $S$, such as an atom, by a macroscopic measuring instrument, ${\cal I}$. 
In Von Neumann\rq s [1] {\it phenomenological} picture, the $S-{\cal I}$  coupling  
leads to two essential effects. Firstly it converts a pure state of $S$, as given by a linear 
combination ${\sum}_{r=1}^{n}c_{r}u_{r}$ of its orthonormal eigenstates $u_{r}$, 
into a statistical mixture of these states for which ${\vert}c_{r}{\vert}^{2}$ is the 
probability of finding this system in the state $u_{r}$: this is the phenomenon often 
termed the \lq reduction of the wave packet\rq. Secondly, it sends a certain set of 
classical, i.e. intercommuting, macroscopic variables $M$ of  ${\cal I}$ to values, 
indicated by pointers,  that specify the state $u_{r}$ of $S$ that is actually realised. The 
problem of the quantum theory of this process is to characterise the properties of the 
macroscopic observables $M$ and the dynamics of the composite $S_{c}=(S+{\cal I})$ 
that lead to these two effects. Our objective here is to treat this problem on the basis of 
the model for which $S_{c}$ is a strictly conservative finite quantum system, whose 
dynamics is governed by its many-particle Schroedinger equation; and our main result is 
that this model does indeed contain the structures required for the resolution of this 
problem. This result provides mathematical justification for the heuristic arguments of 
Van Kampen [2], which led to essentially the same conclusion. It also establishes that 
there is no need to base quantum measurement theory on the model, advocated by some 
authors [3-7], in which $S_{c}$ is a dissipative system, as a result of {\it either} its 
interaction with the \lq rest of the Universe\rq\  [3-6] {\it or} a certain postulated 
nonlinear modification of its Schroedinger equation that leads to a classical deterministic 
evolution of its macroscopic observables [7].
\vskip 0.2cm
As regards the main requirements of a satisfactory theory of the measurement process, it 
is clear from the works of Bohr [8], Jauch [9] and Van Kampen [2] that such a theory 
demands both a characterisation of the macroscopicality of the observables $M$ and an 
amplification property of the $S-{\cal I}$ coupling whereby different microstates of $S$ 
give rise to macroscopically different states of ${\cal I}$. Evidently, this implies that the 
initial state in which ${\cal I}$ is prepared must be unstable against microscopic changes 
in the state of $S$. On the other hand, as emphasised by Whitten-Wolfe and Emch [10, 
11],  the correspondence between the microstate of $S$ and the eventual observed 
macrostate of ${\cal I}$ must be stable against macroscopically small changes in the 
initial state of this instrument, of the kind that are inevitable in experimental procedures. 
Thus, the initial state of ${\cal I}$ must be {\it metastable} by virtue of this combination 
of stability and instability properties. 
\vskip 0.2cm   
There are basically two ways of  characterising the macroscopicality of the observables 
$M$ of ${\cal I}$. The first is to represent this instrument as a large but finite $N$-
particle system for which $N$ is extremely large, e.g. of the order of $10^{24}$. $M$ is 
then represented according to the scheme of Van Kampen [12] and Emch [13] as an 
intercommuting set of observables, which typically are coarse-grained extensive 
conserved variables of  parts or of the whole of  this instrument. The simultaneous 
eigenspaces of these observables then correspond to classical \lq phase cells\rq\ , which 
represent the possible positions of the pointers of ${\cal I}$. Moreover, for suitably 
coarse-grained macroscopic observables $M$, the dimensionality of each of these cells is 
astronomically large [2, 12] since,  by Boltzmann\rq s formula, it is just the exponential 
of the entropy of the macrostate that it represents, and thus it increases exponentially with 
$N$.
\vskip 0.2cm
The second way of characterising the macroscopicality of $M$ is to idealise the 
instrument ${\cal I}$ as an infinitely extended system of particles, with finite number 
density, and to take $M$ to be a set of global intensive observables, which necessarily 
intercommute [14, 10, 11]. This corresponds to the picture employed in the statistical 
mechanical description of large systems in the thermodynamic limit [15-17], and it has 
the merit of sharply distinguishing between macroscopically different states, since 
different values of $M$ correspond to disjoint primary representations of the observables. 
Moreover, in the treatments of the measurement problem based on this idealisation, the 
models of Hepp [14] and Whitten-Wolfe and Emch [10, 11] do indeed exhibit the 
required reduction of the wave-packet and the one-to-one correspondence between the 
pointer position of ${\cal I}$  and the resultant state of $S$; and these results are stable 
against all localised perturbations of the initial state of ${\cal I}$. On the debit side, 
however, Hepp\rq s model requires an infinite time for the measurement to be effected 
(cf. Bell [18]), while although that of Whitten-Wolfe and Emch achieves its 
measurements in finite times, it does so only by dint of a physically unnatural, globally 
extended $S-{\cal I}$ interaction. In view of these observations, it appears to be 
worthwhile to explore the mathematical structure of the measuring process on the basis of 
the model for which ${\cal I}$ is large but finite, with the aim of obtaining conditions 
under which it yields the essential results obtained for the infinite model instrument, but 
with a finite realistic observational time. Evidently, this  requires rigorous control of any 
approximations that arise as a result of the finiteness of $N$.
\vskip 0.2cm
The object of this article is to investigate the mathematical structure of the measurement 
process by means of a dynamical treatment of the generic model of the composite, 
$S_{c}$, of  a microsystem, $S$, and a macroscopic, but finite, measuring instrument 
${\cal I}$. Our treatment of this model is designed to obtain conditions on the $S-{\cal 
I}$ coupling that lead to both the reduction of the wave-packet and the required 
correspondence between the reading of the instrument\rq s pointer and the resultant state 
of $S$.  As in the works [2] , [8-11]  and [14], we avoid the assumption of Von Neumann 
[1] and Wigner [19] that the observation of  the pointer of ${\cal I}$ requires another 
measuring instrument,  ${\cal I}_{2}$, which in turn requires yet another instrument, and 
so on, in such a way that the whole process involves an infinite regression ending up in 
the observer\rq s brain! Instead, we assume that the measurement  process ends with the 
reading of the pointers that evaluate the  macrovariables $M$ of ${\cal I}$. This carries 
the implicit assumption that the dynamics of these variables is sufficiently robust to 
ensure that the act of reading the pointers has negligible effect on their positions. In this 
sense, the macroscopic variables of ${\cal I}$ behaves radically differently from the 
observables of $S$, since the states of  \lq small\rq\ quantum microsystems are 
susceptible to drastic changes as a result of microscopic disturbances. A further crucial 
property of  ${\cal I}$ is that, as pointed out above, the dimensionality of each of its 
macroscopic phase cells is of astronomically large dimensionality, which increases 
exponentially with $N$; and, by a similar argument, the same is true for its density of 
energy eigenstates We shall see that the enormous phase cells of the finite instrument 
${\cal I}$ play the essential role of the disjoint representation spaces of the infinite one 
and consequently that the finite model possesses all the positive properties of the infinite 
one, with the bonus that it achieves its measurements in finite, realistic times. 
Furthermore, the enormity of its density of energy eigenstates ensures that the periods of 
its Poincare recurrences are astronomically long. We can therefore discount these 
recurrences by restricting our treatment of the dynamics to finite intervals of much 
shorter duration.
\vskip 0.2cm
Turning now to the formulation of the measurement process, we assume, in a standard 
way, that the observables of $S$ and the macroscopic ones, $M$, of  ${\cal I}$, on which 
measurements are performed, generate $W^{\star}$-algebras ${\cal A}$ and ${\cal M}$, 
respectively, the latter being abelian. The process is expressed in terms of the state on the 
algebra ${\cal A}{\otimes}{\cal M}$ that results from the evolution of $S_{c}$ from an 
initial state obtained by independent preparations of $S$  and ${\cal I}$. The resultant 
evolved state of $S_{c}$ then determines the expectation values of the observables, $A$, 
of  $S$ and their conditional expectation values, $E(A{\vert}{\cal M})$, given the 
$M$\rq s. Thus it determines the probabilistic state, ${\rho}$, of $S$ {\it before} those 
macroscopic variables are  measured and its {\it subsequent} state, as given by the form 
of $E(.{\vert}{\cal M})$, following the measurement.
\vskip 0.2cm
On relating the state ${\rho}$ and the conditional expectation functional $E(.{\vert}{\cal 
M})$ to the $S-{\cal I}$ interaction, we find that the mathematical model yields two 
classes of effective instruments ${\cal I}$, though from the empirical standpoint these 
classes are essentially equivalent. The first class of instruments comprises those for 
which the wave packet of $S$ is reduced according to Von Neumann\rq s prescription 
and the correspondence between the pointer position of ${\cal I}$ and the microstate of 
$S$ is strictly one-to-one. The second class of instruments comprises those for which this 
result arises with overwhelming probability, for large $N$, rather than with absolute 
certainty. Thus, in this case, if the result of a measurement is interpreted on the basis of 
an assumption of a perfect correspondence between the microstate of $S$ and the 
macrostate of ${\cal I}$,  then there is a miniscule possibility that the pointer position 
will correspond to a state of $S$ quite different from (in general orthogonal to) the 
indicated one. We term the instruments of the first class {\it ideal} and those of the 
second class {\it normal}. As support for this classification of instruments, we show that, 
in the case of  a finite version of the Coleman-Hepp model [14], the instrument is 
generically normal, though it is ideal for certain special values of its parameters. 
Furthermore, in the former case, the odds against the pointer indicating the \lq wrong\rq\ 
state of $S$ increase exponentially with $N$.
\vskip 0.2cm
We present our mathematical treatment of the measurement process as follows. In 
Section 2, we formulate the generic model of the composite quantum system $S_{c}$, 
employing the phase cell representation of Van Kampen [12] and Emch [13] for the 
description of the macroscopic observables of ${\cal I}$.  In particular, we formulate the 
time-dependent expectation values of the observables, $A$, of $S$ and of the 
macroscopic ones, $M$,  of ${\cal I}$, as well as the conditional expectation value of  
$A$, given ${\cal M}$, subject to the assumption that $S$ and ${\cal I}$ are 
independently prepared and then coupled together at time $t=0$. In Section 3, we 
formulate the conditions on the dynamics of the model under which the measuring 
instrument ${\cal I}$ is ideal or normal, in the sense described above. In Section 4, we 
show that the general scheme of Sections 2 and 3 is fully realized by the finite version of 
the Coleman-Hepp model [14]. There we show that the instrument ${\cal I}$ for this 
model is generically normal, though for certain special values of its parameters it is ideal. 
Moreover, we show that these results are stable under localised perturbations of the initial 
state of ${\cal I}$, and even under global ones that correspond to small changes in the 
values of intensive thermodynamical variables (e.g. temperature, polarisation) of that 
state. Here we take the generic prevalence of normality of ${\cal I}$, for this model, to be 
an indication that real quantum measuring instruments are generally normal rather than 
ideal. We conclude, in Section 5, with a brief resume of the picture presented here.and a 
suggestion about a possible further development in the physics of the quantum 
measutrement process.
\vskip 0.5cm
\centerline {\bf 2. The Generic Model.} 
\vskip 0.3cm
We assume that the algebras of observables, ${\cal A}$ and ${\cal B}$, of the 
microsystem $S$ and the instrument  ${\cal I}$, are those of the bounded operators in 
separable Hilbert spaces  ${\cal H}$  and ${\cal K}$, respectively. Correspondingly, the 
states of  these systems are represented by the density matrices in the respective spaces. 
The density matrices for the pure states are then the one-dimensional projectors. For 
simplicity we assume that ${\cal H}$ is of finite dimensionality $n$.
\vskip 0.2cm
We assume that the coupled composite $S_{c}:=(S+{\cal I})$ is a conservative system, 
whose Hamiltonian operator $H_{c}$, in ${\cal H}{\otimes}{\cal K}$, takes the form
$$H_{c}=H{\otimes}I_{\cal K}+I_{\cal H}{\otimes}K+V,\eqno(2.1)$$
where $H$ and $K$ are the Hamiltonians of $S$ and ${\cal I}$, respectively, and  $V$ is 
the $S-{\cal I}$ interaction. Thus, the dynamics of $S_{c}$ is given by the one-
parameter group $U_{c}$ of  unitary transformations of ${\cal H}{\otimes}{\cal K}$ 
generated by $iH_{c}$, i.e. 
$$U_{c}(t)={\rm exp}(iH_{c}t) \ {\forall} \ t{\in}{\bf R}.\eqno(2.2)$$
We assume that the the systems $S$ and ${\cal I}$ are prepared, independently of one 
another, in their initial states represented by density matrices ${\omega}$ and 
${\Omega}$, respectively, and then coupled together at time $t=0$. Thus the initial state 
of the composite $S_{c}$ is given by the density matrix ${\omega}{\otimes}{\Omega}$ 
in ${\cal H}_{c}:={\cal H}{\otimes}{\cal K}$. Further, we assume that the initial state 
of $S$ is pure, and thus that ${\omega}$ is the projection operator $P({\psi})$ for a 
vector ${\psi}$ in ${\cal H}$. The initial state of $S_{c}$ is then
$${\Phi}=P({\psi}){\otimes}{\Omega}.\eqno(2.3)$$
Since ${\cal H}$ is $n$-dimensional, we may take as its basis a complete orthonormal set 
of eigenvectors, $(u_{1},. \  .,u_{n})$, of $H$ .  Hence, the initial state vector ${\psi}$ of 
$S$ is given by a linear combination of these vectors, i.e.
$${\psi}={\sum}_{r=1}^{n}c_{r}u_{r},\eqno(2.4)$$
where
$${\sum}_{r=1}^{n}{\vert}c_{r}{\vert}^{2}=1;\eqno(2.5)$$
while the action of $H$ on $u_{r}$ is given by the equation
$$Hu_{r}={\epsilon}_{r}u_{r},\eqno(2.6)$$
where ${\epsilon}_{r}$ is the corresponding eigenvalue of this operator.
\vskip 0.2cm
We assume that the instrument ${\cal I}$ is designed to perform measurements of the 
first kind (cf. Jauch [9]), whereby the $S-{\cal I}$ coupling does not induce transitions 
between the eigenstates ${\lbrace}u_{r}{\rbrace}$ of $S$. This signifies that the 
interaction $V$ takes the form
$$V={\sum}_{r=1}^{n}P(u_{r}){\otimes}V_{r},$$
where $P(u_{r})$ is the projection operator for $u_{r}$ and the $V_{r}$\rq s are 
observables of ${\cal I}$. Hence, by Eq. (2.1), the Hamiltonian of the composite system 
$S_{c}$ is
$$H_{c}={\sum}_{r=1}^{n}P(u_{r}){\otimes}K_{r},\eqno(2.7)$$
where
$$K_{r}=K+V_{r}+{\epsilon}_{r}I_{\cal K}.\eqno(2.8)$$
Consequently, by Eqs. (2.2) and (2.7), the dynamical group $U_{c}$ is given by the 
formula
$$U_{c}(t)={\rm exp}(iH_{c}t)={\sum}_{r=1}^{n}P(u_{r}){\otimes}U_{r}(t),
\eqno(2.9)$$
where
$$U_{r}(t)={\rm exp}(iK_{r}t).\eqno(2.10)$$
Consequently, since the evolute at time $t \ ({\geq}0)$ of the initial state ${\Phi}$ of 
$S_{c}$ is $U_{c}^{\star}(t){\Phi}U_{c}(t):={\Phi}(t)$,  it follows from Eqs.(2.3), (2.4) 
and (2.10) that
$${\Phi}(t)={\sum}_{r,s=1}^{n}{\overline c}_{r}c_{s}P_{r,s}
{\otimes}{\Omega}_{r,s}(t),\eqno(2.11)$$
where $P_{r,s}$ is the operator in ${\cal H}$ defined by the equation
$$P_{r,s}f=(u_{s},f)u_{r} \ {\forall} \ f{\in}{\cal H}\eqno(2.12)$$
and  
$${\Omega}_{r,s}(t)=U_{r}^{\star}(t){\Omega}U_{s}(t).\eqno(2.13)$$
\vskip 0.3cm
{\bf  2.1. The Macroscopic Observables of ${\cal I}$.} We assume that these conform to 
the following scheme, due to Van Kampen [12] and Emch [13].
\vskip 0.2cm\noindent
(1) They are intercommuting observables, which typically  are coarse grained extensive 
conserved variables of parts or of the whole of the system ${\cal I}$. The algebra, ${\cal 
M}$, of these observables is therefore an abelian subalgebra of  the full algebra, ${\cal 
B}$,  of bounded observables of ${\cal I}$. For simplicity, we assume that ${\cal M}$ is 
finitely generated and thus that it consists of  the linear combinations of a finite set of 
orthogonal projectors ${\lbrace}{\Pi}_{\alpha}{\vert}{\alpha}=1,2. \ .,{\nu}{\rbrace}$ 
that span the space ${\cal K}$. It follows from these specifications that
$${\Pi}_{\alpha}{\Pi}_{\beta}={\Pi}_{\alpha}{\delta}_{{\alpha}{\beta}},\eqno(2.14)$$
$${\sum}_{{\alpha}=1}^{\nu}{\Pi}_{\alpha}=I_{\cal K}\eqno(2.15)$$
and that any element, $M$, of ${\cal M}$ takes the form 
$$M={\sum}_{{\alpha}=1}^{\nu}M_{\alpha}{\Pi}_{\alpha},\eqno(2.16)$$
where the $M_{\alpha}$\rq s are constants.The subspaces ${\lbrace}{\cal K}_{\alpha}:=
{\Pi}_{\alpha}{\cal K}{\rbrace}$ of ${\cal K}$ correspond to classical  phase cells. 
Each such cell then represents a macrostate of ${\cal I}$, and is identified by the position 
of a pointer (or set of pointers) in a measurement process. 
\vskip 0.2cm\noindent
(2) As noted in Section 1, the dimensionality of each cell ${\cal K}_{\alpha}$ is 
astronomically large, since it is given essentially by the exponential of the entropy 
function of the macro-observables and thus grows exponentially with $N$. The largeness 
of the phase cells is closely connected to the robustness of the macroscopic measurement.
\vskip 0.2cm
Note here that these properties of ${\cal M}$ are just general ones of macroscopic 
observables and do not depend on ${\cal I}$ being a measuring instrument for the system 
$S$. The coordination of these properties with those of $S$ that are pertinent to the 
measuring process will be treated in Section 3.
\vskip 0.3cm 
{\bf 2.2. Expectation and Conditional Expectation Values of Observables.} The 
observables of $S_{c}$ with which we shall be concerned are just the self-adjoint 
elements of ${\cal A}{\otimes}{\cal M}$. Their expectation values for the time-
dependent state ${\Phi}(t)$ are given by the formula
$$E\bigl(A{\otimes}M\bigr)={\rm Tr}\bigl({\Phi}(t)[A{\otimes}M]\bigr) \ {\forall} \ 
A{\in}
{\cal A}, \ M{\in}{\cal M},\eqno(2.17)$$
In particular, the expectation values of the observables of $S$ and the macroscopic ones 
of ${\cal I}$ are given by the equations
$$E(A)=E(A{\otimes}I_{\cal K})\eqno(2.18)$$
and
$$ E(M)=E(I_{\cal H}{\otimes}M) ,\eqno(2.19)$$
respectively. Further, in view of the abelian character of ${\cal M}$, the expectation 
functional $E$ is compatible with a unique conditional expectation functional on ${\cal 
A}$ with respect to ${\cal M}$, as the following argument shows. Such a conditional 
expectation is a linear mapping $E(.{\vert}{\cal M})$ of ${\cal A}$ into ${\cal M}$ that 
preserves positivity and normalisation and satisfies the condition
$$E\bigl(E(A{\vert}{\cal M})M\bigr)=E(A{\otimes}M) \ {\forall} \ A{\in}{\cal A}, \ 
M{\in}{\cal M}.\eqno(2.20)$$ 
Therefore since, by linearity and Eq. (2.16), $E(.{\vert}{\cal M})$ must take the form
$$E(A{\vert}{\cal M})={\sum}_{\alpha}{\omega}_{\alpha}(A){\Pi}_{\alpha},
\eqno(2.21)$$
where the ${\omega}_{\alpha}$\rq s are linear functionals on ${\cal A}$, it follows from 
Eq. (2.20) that
$${\omega}_{\alpha}(A)E({\Pi}_{\alpha})=E(A{\otimes}{\Pi}_{\alpha})$$
and consequently, by Eq. (2.21),
$$E(A{\vert}{\cal M})={\sum}_{\alpha}^{\prime}E(A{\otimes}{\Pi}_{\alpha})
{\Pi}_{\alpha}/ E({\Pi}_{\alpha})  \  {\forall} \ A{\in}{\cal A},\eqno(2.22)$$
where the prime over ${\Sigma}$ indicates that summation is confined to the 
${\alpha}$\rq s for which $E({\Pi}_{\alpha})$ does not vanish. In view of Eq. (2.15),  
this formula for $E(.{\vert}{\cal M})$ meets the requirements of positivity and 
normalisation. 
\vskip 0.3cm
{\bf 2.3.  Properties of the Expectation Functional $E$.} By Eqs. (2.11)-(2.13), (2.16) 
and (2.17),
$$E\bigl(A{\otimes}M\bigr)={\sum}_{r,s=1}^{n}{\sum}_{{\alpha}=1}^{\nu}
{\overline c}_{r}c_{s}(u_{r},Au_{s})M_{\alpha}F_{r,s;{\alpha}},\eqno(2.23)$$
where
$$F_{r,s;{\alpha}}={\rm Tr}\bigl({\Omega}_{r,s}(t){\Pi}_{\alpha}\bigr).\eqno(2.24)$$
Key properties of $F_{r,s:{\alpha}}$, which follows from Eqns. (2.13), (2.15) and (2.24) 
are that
$${\sum}_{{\alpha}=1}^{\nu}F_{r,r;{\alpha}}=1,\eqno(2.25)$$
$$1{\geq}F_{r,r;{\alpha}}{\geq}0\eqno(2.26)$$
and
$$F_{r,s:{\alpha}}={\overline F}_{s,r:{\alpha}},\eqno(2.27)$$
where the bar over $F$ on the r.h.s. indicates complex conjugation. It also follows from 
those formulae that, for 
$z_{1},. \ .,z_{n}{\in}{\bf C}$, the sesquilinear form ${\sum}_{r,s=1}^{n}
{\overline z}_{r}z_{s}F_{r,s;{\alpha}}$ is positive, and hence
$$F_{r,r;{\alpha}}F_{s,s;{\alpha}}{\geq}{\vert}F_{r,s;{\alpha}}{\vert}^{2}.
\eqno(2.28)$$
\vskip 0.5cm
\centerline {\bf 3. The Measurement Process}
\vskip 0.3cm
As noted in Section 2, a pointer reading of ${\cal I}$ serves to identify the phase cells 
${\cal K}_{\alpha}$ that represents its macrostate. Eq. (2.22) therefore signifies that the 
expectation values of the observables of $S$ following that measurement is given by the 
formula
$$E(A{\vert}{\cal K}_{\alpha}):= E(A{\otimes}{\Pi}_{\alpha})/E({\Pi}_{\alpha}).
\eqno(3.1)$$
Now, in order that the pointer reading specifies the eigenstate of $S$, we require a one-
to-one correspondence between the  phase cells ${\cal K}_{\alpha}$ and the eigenstates 
$u_{r}$ of $S$. Accordingly, we assume that, for an instrument designed to identify the 
microstate of $S$, the number of these phase cells is just the number of  the eigenstates 
$u_{r}$ of $S$, namely $n$.
\vskip 0.3cm
{\bf 3.1. The Ideal Instruments.} We term the instrument ${\cal I}$ {\it ideal} if  there is 
a one-to-one correspondence between the pointer reading ${\alpha}$ and the eigenstate 
$u_{r}$ of $S$, on a realistic observational time scale. Thus ${\cal I}$ is ideal if, for 
times $t$ greater than some critical value, ${\tau}$, and less, in order of magnitude, than 
the Poincare' recurrence times, the following conditions are fulfilled.
\vskip 0.2cm\noindent
(I.1) the time-dependent state ${\Omega}_{r,r}(t)$ of ${\cal I}$, that arises in 
conjunction with the state $u_{r}$ of $S$ in the formula Eqs. (2.13), lies in one of the 
subspaces ${\cal K}_{\alpha}$ of ${\cal K}$; 
\vskip 0.2cm\noindent
(I.2) the correspondence between $r$ and ${\alpha}$ here is one-to-one, i.e. 
${\alpha}=a(r)$, where $a$ is an invertible transformation of the point set ${\lbrace}1,2,. 
\ .,n{\rbrace}$; and
\vskip 0.2cm\noindent
(I.3) this correspondence is stable with respect to perturbations of the initial state 
${\Omega}$ of ${\cal I}$ that are localised, in the sense that each of them leaves this 
state unchanged outside some region contained in a ball of volume $O(1)$ with respect to 
$N$. 
\vskip 0.2cm\noindent
The conditions (I.1) and (I.2) signify that, for times $t$ in the range specified there,
$${\rm Tr}\bigl({\Omega}_{r,r}(t){\Pi}_{\alpha}\bigr)={\delta}_{a(r),{\alpha}},$$
i.e., by Eq. (2.24),
$$F_{r,r:{\alpha}}={\delta}_{a(r),{\alpha}}.\eqno(3.2)$$
Moreover, it follows from Eqs. (2.25) and (2.26), together with the invertibility of the 
function $a$, that Eq. (3.2) not only implies but is actually equivalent to the condition
$$F_{r,r;a(r)}=1.\eqno(3.2)^{\prime}$$
Further, by Eqs. (2.28) and (3.2) and the invertibility of $a$,
$$F_{r,s;{\alpha}} =0 \ {\rm for} \ r{\neq}s.\eqno(3.3)$$
Consequently, by Eqs. (2.23), (3.2) and (3.3), 
$$E(A{\otimes}M)={\sum}_{r=1}^{n}w_{a(r))}M_{a(r)}
(u_{r},Au_{r}),\eqno(3.4)$$
where
$$w_{a(r)}={\vert}c_{r}{\vert}^{2}.\eqno(3.5)$$
Hence, by  Eqs. (3.1) and (3.4) and the invertibility of $a$,
$$E({\Pi}_{\alpha})=w_{\alpha},\eqno(3.6)$$
$$E(A)={\sum}_{r=1}^{n}w_{a(r)}
(u_{r},Au_{r})\eqno(3.7)$$
and
$$E(A{\vert}{\cal K}_{a(r)})= (u_{r},Au_{r}).\eqno(3.8 )$$
\vskip 0.2cm
Eqs. (3.6) and (3.7) signify that, {\it before} the pointer position is read, $w_{\alpha}$ is 
the probability that the reading is ${\alpha}$ and the state of $S$ is given by the density 
matrix 
$${\rho}={\sum}_{r=1}^{n}w_{a(r)}P(u_{r}),$$
i.e., by Eq. (3.5),
$${\rho}={\sum}_{r=1}^{n}{\vert}c_{r}{\vert}^{2}P(u_{r}).\eqno(3.9)$$
Thus we have a reduction of the wave packet, as given by the  transition from the pure 
state ${\psi} \ (={\sum}_{r=1}^{n})$ to this mixed state ${\rho}$. 
\vskip 0.2cm
According to the standard probabilisitic interpretation of quantum mechanics, Eq. (3.9) 
specifies the state of $S$ just prior to the reading of the pointers, whereas Eq. (3.8) serves 
to specify its state following that reading. Thus, by Eq. (3.9), ${\vert}c_{r}{\vert}^{2}$ 
is the probability that the pointer reading will yield the result that $u_{r}$ is the state of 
$S$; while Eq. (3.8) signifies that, following a reading that yields the result  that 
${\alpha}=a(r)$, the state of $S$ is $u_{r}$. In the standard picture of quantum theory, 
there is no causality principle that determines which of the states $u_{r}$ will be found. 
\vskip 0.3cm
{\bf Comments.} (1)  As shown above, the property (3.2) ensures that ${\cal I}$ enjoys 
all the essential properties of a measuring instrument since it implies both the reduction 
of the wave-packet and the one-to-one correspondence between the pointer position and 
the microstate of  $S$. On the other hand, the property (3.3), which ensures the reduction 
of the wave-packet, does not imply Eq. (3.2) and therefore does not, of itself, imply that 
${\cal I}$ serves as a measuring instrument
\vskip 0.2cm
(2)  The property (3.3) signifies that the $S-{\cal I}$ coupling removes the interference 
between the different components $u_{r}$ of the pure state ${\psi}$ and thus represents 
a {\it complete decoherence} effect. To see how this is related to the structure of a typical 
phase cell ${\cal K}_{\alpha}$, we introduce a complete orthonormal basis 
${\lbrace}{\theta}_{{\alpha},{\lambda}}{\rbrace}$ of this cell, where the index 
${\lambda}$ runs from $1$ to ${\rm dim}({\cal K}_{\alpha})$, the dimensionality of 
${\cal K}_{\alpha}$. We then infer from Eqs. (2.13) and (2.24) that
$$F_{r,s;{\alpha}}={\sum}_{{\lambda}=1}^{{\rm dim}({\cal K}_{\alpha})}
\bigl(U_{r}(t){\theta}_{{\alpha},{\lambda}},{\Omega} 
U_{s}(t){\theta}_{{\alpha},{\lambda}}\bigr),$$
Hence, as $iK_{r}$ is the generator of $U_{r}$, this equation signifies that the 
decoherence arises from the aggregated destructive interference of the evolutes of the 
vectors ${\theta}_{{\alpha},{\lambda}}$ generated by the different Hamiltonians 
$K_{r}$ and $K_{s}$.  This picture of decoherence corresponds to that assumed by Van 
Kampen [2].
\vskip 0.3cm
{\bf 3.2. Normal Measuring Instruments.} We term the instrument ${\cal I}$ 
{\it  normal}\footnote*{We conjecture that the behaviour of real instruments is generally 
normal in the sense specified here and thus that the use of this adjective is approriate. 
Some support for this conjecture is provided by the results of Section 4 for the Coleman-
Hepp model.} if  the following conditions are fulfilled.
\vskip 0.2cm\noindent
(N.1) A weaker form of the ideality condition (3.2),  or equivalently (3.2)$^{\prime}$, 
prevails, to the effect that the difference between the two sides of the latter formula is 
negligibly small, i.e., noting Eq. (2.25), that 
$$0<1-F_{r,r;a(r)}<{\eta}(N),\eqno(3.10)$$  
where, for large $N, \ {\eta}(N)$  is miniscule by comparison with unity: in the case of 
the finite version of the Coleman-Hepp model treated in Section 4, it is ${\exp}(-cN)$, 
where $c$ is a fixed positive constant of the order of unity. We note that, by Eq. (2.25) 
and the positivity of ${\Pi}_{\alpha}$, the condition (3.10) is equivalent to the inequality
$$0<{\sum}_{r{\neq}a^{-1}({\alpha})}F_{r,r;{\alpha}}<{\eta}(N).
\eqno(3.10)^{\prime}$$
Further, it follows from Eqs. (2.28), (3.10) and (3.10)$^{\prime}$ that 
$${\vert}F_{r,s;{\alpha}}{\vert}<{\eta}(N)^{1/2} \ 
{\rm for} \ r{\neq}s,\eqno(3.11)$$
which is evidently a {\it decoherence condition}, being a slightly weakened version of 
the complete one given by Eq. (3.3).
\vskip 0.2cm\noindent 
(N.2) This condition (N.1) is stable under localised modifications of the initial state 
${\Omega}$ of ${\cal I}$. This stability condition may even be strengthened to include 
global perturbations of ${\Omega}$ corresponding to small changes in its intensive 
thermodynamic parameters (cf. the treatment of the Coleman-Hepp model in Section 4).
\vskip 0.2cm 
It follows now from Eq. (3.11) that the replacement of the ideal condition (3.2) by the 
normal one (3.10)  leads to modifications of the order ${\eta}(N)^{1/2}$ to the formula 
(3.4) and its consequences. In particular, it implies that a pointer reading ${\alpha}$ 
signifies that it is overwhelmingly probable, but not absolutely certain, that the state of 
$S$ is 
$u_{a^{-1}({\alpha})}$, as the following argument shows. Suppose that the initial state 
of $S$ is $u_{r}$. Then, by Eq. (2.11), the state of $S_{c}$ at time $t$ is 
$P(u_{r}){\otimes}{\Omega}_{r,r}(t)$; and by Eqs. (3.10)$^{\prime}$, there is a 
probability of the order of ${\eta}(N)$ that the pointer reading is given by a value 
${\alpha}$, different from $a(r)$, of the indicator parameter of ${\cal I}$. In the freak 
case that this possibility is realised, this would mean that the state $u_{r}$ of $S$ led to a 
pointer reading ${\alpha}{\neq}a(r)$. Hence, in this case, any inference to the effect that 
a pointer reading ${\alpha}$ signified that the state of $S$ was $u_{a^{-1}({\alpha})}$ 
would be invalid. 
\vskip 0.3cm
{\bf Comments.} The scheme proposed here admits two kinds of measuring instruments, 
namely the ideal and the normal ones. The former fulfill perfectly the demands for the 
reduction of the wave-packet and the one-to-one correspondence between the pointer 
reading of the measuring instrument and the eigenstate of the observed microsystem. On 
the other hand, in the case of a normal instrument, there is just a minuscule possibility 
that the pointer reading might correspond to the \lq wrong\rq\ eigenstate of the 
microsystem. However, as the odds against such an eventuality are overwhelming, the 
distinction between the two kinds of instruments is essentially mathematical rather than 
observational.
\vskip 0.5cm
\centerline {\bf 4. The Finite Coleman-Hepp Model.} 
\vskip 0.3cm
This model is a caricature of an electron that interacts with a finite spin chain that serves 
to measure the electronic spin [14]. In order to fit this model into the scheme of the 
previous Sections, we regard the electron, ${\cal P}$, as the composite of two entities, 
namely its spin, ${\cal P}_{1}$, and its orbital component, ${\cal P}_{2}$. We then take 
the system $S$ to be just ${\cal P}_{1}$ and the instrument ${\cal I}$ to be the 
composite of ${\cal P}_{2}$ and the chain ${\cal C}$. Thus, we build the model of 
$S_{c}=(S+{\cal I})$ from its components in the following way.
\vskip 0.3cm
{\bf 4.1. The System $S={\cal P}_{1}$.} This is just a single Pauli spin. Thus, its state 
space is ${\cal H}={\bf C}^{2}$ and its three-component spin observable is given by the 
Pauli matrices $(s_{x},s_{y},s_{z})$. We denote by $u_{\pm}$ the eigenvectors of 
$s_{z}$ whose eigenvalues are ${\pm}1$, respectively. These vectors then form a basis 
in ${\cal H}$. We denote their projection operators by $P_{\pm}$, respectively.
\vskip 0.3cm
{\bf 4.2. The System ${\cal I}=({\cal P}_{2}+{\cal C})$.} We assume that ${\cal P}$ 
moves along, or parallel to, the axis $Ox$ and thus that the state space of ${\cal P}_{2}$ 
is ${\tilde {\cal K}}:=L^{2}({\bf R})$. We assume that ${\cal C}$ is a chain of Pauli 
spins located at the sites $(1,2,. \  .,2L+1)$, of  $Ox$, where $L$ is a positive integer. 
Thus, the state space of ${\cal C}$ is ${\hat {\cal K}}:=({\bf C}^{2})^{(2L+1)}$, and 
therefore that of ${\cal I}$ is ${\cal K}={\tilde {\cal K}}{\otimes}{\hat {\cal K}}$. 
\vskip 0.2cm
The spin at the site $n$ of ${\cal C}$ is represented by Pauli matrices 
$({\sigma}_{n,x},{\sigma}_{n,y},{\sigma}_{n,z})$ that act on the $n$\rq th 
${\bf C}^{2}$ component of ${\hat {\cal K}}$. Thus, they may be canonically identified 
with operators in  ${\hat {\cal K}}$ that satisfy the standard Pauli relations
$${\sigma}_{n,x}^{2}={\sigma}_{n,y}^{2}={\sigma}_{n,z}^{2}={\hat I}; \ 
{\sigma}_{n,x}{\sigma}_{n,y}=i{\sigma}_{n,z}, \ {\rm etc},\eqno(4.1)$$
together with the condition that the spins at different sites intercommute.
\vskip 0.2cm
We assume that ${\cal P}_{1}, \ {\cal P}_{2}$ and ${\cal C}$ are independently 
prepared before being coupled together at time $t=0$. Further, we assume that the initial 
states of ${\cal P}_{1}$ and ${\cal P}_{2}$ are the pure ones, represented by vectors 
${\psi}$ and  ${\phi}$ in ${\cal H}$ and ${\tilde {\cal K}}$, respectively, while that of 
${\cal C}$ is given by a density matrix ${\hat {\Omega}}$, in  ${\hat {\cal K}}$, whose 
form will be specified below, by Eqs. (4.3) and (4.4).Thus, the initial state of ${\cal  I}$ 
is
$${\Omega}=P({\phi}){\otimes}{\hat {\Omega}},\eqno(4.2)$$
where $P({\phi})$ is the projection operator for ${\phi}$. We assume that ${\phi}$ has 
support in a finite interval $[c,d]$ and that ${\hat {\Omega}}$ takes the form
$${\hat {\Omega}}={\otimes}_{n=1}^{2L+1}{\hat {\omega}}_{n},\eqno(4.3)$$
where ${\hat {\omega}}_{n}$, the initial state of the $n$\rq th spin of ${\cal C}$, is give 
by the formula
$${\hat {\omega}}_{n}={1\over 2}(I_{n}+m{\sigma}_{n,z}),\eqno(4.4)$$
where $0<m{\leq}1$. Thus, asuming that there are no interactions between the spins of 
${\cal C}, \ {\hat {\Omega}}$ is the equilibrium state obtained by subjecting this chain 
to a certain temperature-dependent magnetic field, directed along $Oz$. $m$ is then the 
magnitude of the resultant polarisation of this chain. One sees immediately from Eqs. 
(4.3) and (4.4) that ${\hat {\Omega}}$ is a pure state if $m=1$: otherwise it is mixed. 
\vskip 0.3cm
{\bf 4.3. The Dynamics.} Following Hepp [14], we assume that the Hamiltonian for the 
composite system $S_{c}$ is
$$H_{c}=I_{\cal H}{\otimes}p{\otimes}I_{\hat {\cal K}}+
P_{-}{\otimes}{\sum}_{n=1}^{2L+1}V(x-n){\otimes}{\sigma}_{n,x},\eqno(4.5)$$
where $p$ and $V$ are the differential and multiplicative operators in $L^{2}({\bf R}) \ 
(={\tilde {\cal K}})$ that transform $f(x)$ to $-i{\hbar}df(x)/dx$ and $V(x)f(x)$, 
respectively, and $V$ is a bounded, real valued function on ${\bf R}$ with support in a 
finite interval $[a,b]$. Thus, in the notation of Eq. (2.8), but with $r$ taking just the 
values ${\pm}$,
$$K_{+}=p{\otimes}I_{\hat {\cal K}}  \ {\rm and} \ K_{-}=
p{\otimes}I_{\hat {\cal K}}+
{\sum}_{n=1}^{2L+1}V(x-n){\otimes}{\sigma}_{n,x}.\eqno(4.6)$$
The assumption here that the Hamiltonian for the free orbital motion of ${\cal P}$ is 
linear rather than quadratic in $p$ serves to simplify the model by avoiding dispersion of 
the \lq electronic wave packet\rq\ .
\vskip 0.2cm 
The unitary groups $U_{\pm}$ generated by $iK_{\pm}$ are given by the formula
$$U_{\pm}(t) ={\exp}(iK_{\pm}t)\eqno(4.7)$$ 
and  the evolutes of ${\Omega}$ due to the actions of $U_{\pm}(t)$ are 
$${\Omega}_{\pm}(t):=U_{\pm}^{\star}(t){\Omega}U_{\pm}(t).\eqno(4.8)$$
These states are evidently the versions, for this model, of ${\Omega}_{r,r}(t)$, as 
defined by Eq. (2.13),  with the double suffix $(r,r)$ represented by $+$ or $-$. It follows 
now from Eqs. (4.2) and (4.6)-(4.8) that
$${\Omega}_{+}(t)=P({\phi}_{t}){\otimes}{\hat {\Omega}},\eqno(4.9)$$
where
$${\phi}_{t}(x)={\phi}(x+t).\eqno(4.10)$$
As for ${\Omega}_{-}$ it is convenient to formulate its evolution in interaction 
representation, in terms of the unitary operator
$$W(t):=U_{-}(t){\rm exp}(-i[p{\otimes}I_{\hat {\cal K}}]t).\eqno(4.11)$$
Thus, by Eqs. (4.2), (4.8) and (4.11),
$${\Omega}_{-}(t)= {\rm exp}(-i[p{\otimes}I_{\hat {\cal K}}]t)
\bigl(W^{\star}(t)[P({\phi}){\otimes}{\hat {\Omega}}]W(t)\bigr)
{\rm exp}(i[p{\otimes}I_{\hat {\cal K}}]t).\eqno(4.12)$$
By Eqs. (4.6), (4.7) and (4.11),  $W(t)$ satisfies the Dyson integral equation
$$W(t)=I_{\cal K}+i\int_{0}^{t}ds{\sum}_{n=1}^{2L+1}[V(x+s-n){\otimes}
{\sigma}_{n,x}]W(s),$$
whose solution is
 $$W(t)={\rm exp}\bigl(i{\sum}_{n=1}^{2L+1}[F_{n,t}(x){\otimes}
{\sigma}_{n,x}]\bigr),\eqno(4.13)$$
where
$$F_{n,t}(x)=\int_{0}^{t}dsV(x+s-n).\eqno(4.14)$$
Further, since the supports of $V$ and ${\phi}$ are $[a,b]$ and $[c,d]$, respectively, it 
follows from these last two equations that we may replace $F_{n,t}(x)$ by 
$\int_{\bf R}dxV(x)$ when employing Eq. (4.13) in the formula (4.12), provided that 
$$d{\leq}a+1 \ {\rm and} \  t{\geq}{\tau}:=2L+1-b-c.\eqno(4.15)$$
Thus, in this case, $W(t)$ may be replaced there by $I_{{\tilde {\cal K}}}{\otimes}Z$, 
where
$$Z={\rm exp}(iJ{\sum}_{n=1}^{2L+1}{\sigma}_{n,x}){\equiv}
{\otimes}_{n=1}^{2L+1}{\rm exp}(iJ{\sigma}_{n,x})\eqno(4.16)$$
and
$$J=\int_{\bf R}dxV(x).\eqno(4.17)$$
Consequently, under the conditions (4.15), Eq. (4.12) reduces to the form
$${\Omega}_{-}(t)=P({\phi}_{t}){\otimes}Z^{\star}{\hat {\Omega}}Z,$$
where ${\phi}_{t}$ is given by Eq. (4.10). On combining this equation with Eq. (4.9), we 
see that
$${\Omega}_{\pm}(t)=P({\phi}_{t}){\otimes}{\hat {\Omega}}_{\pm},\eqno(4.18),$$
where ${\hat {\Omega}}_{\pm}$ are the {\it time-independent} states given by the 
formulae
$${\hat {\Omega}}_{+}={\hat {\Omega}} \ {\rm and} \ 
{\hat {\Omega}}_{-}=Z^{\star}{\hat {\Omega}}Z.\eqno(4.19)$$
Thus, under the conditions (4.15), the chain ${\cal C}$ takes up the steady states ${\hat 
{\Omega}}_{\pm}$ corresponding to the states $u_{\pm}$ of $S$. It should be noted 
that the critical time ${\tau}$, specified in Eq. (4.15), is essentially the time required for 
the particle ${\cal P}$ to travel from end to end of the chain ${\cal C}$. It is therefore a 
reasonable macroscopic observational time. 
\vskip 0.2cm
Further, by Eqs. (4.1)-(4.4), (4.16) and (4.19), the explict forms of the states 
${\hat {\Omega}}_{\pm}$ are given by the equations
$${\hat {\Omega}}_{+}=2^{-(2L+1)}
{\otimes}_{n=1}^{2L+1}(I_{n}+m{\sigma}_{n,z})\eqno(4.20)$$ 
and 
$${\hat {\Omega}}_{-}=2^{-(2L+1)}{\otimes}_{n=1}^{2L+1}
\bigl(I_{n}+m{\sigma}_{n,z}{\rm cos}(2J)+m{\sigma}_{n,y}{\rm sin}(2J)\bigr).
\eqno(4.21)$$
\vskip 0.3cm
{\bf 4.4. The Macroscopic Phase Cells of  ${\cal I}$.} We take these to be the subspaces 
${\cal K}_{\pm}$ of ${\cal K}$ corresponding to positive and negative polarizations, 
respectively, of the chain ${\cal C}$ along the $Oz$-direction. To formulate these 
subspaces precisely, we first note that the eigenvalues of the total spin of ${\cal C}$ 
along that direction, namely ${\Sigma}_{z}:={\sum}_{n=1}^{(2L+1)}{\sigma}_{z}$, 
are the odd numbers between $-(2L+1)$ and $(2L+1)$. We define ${\hat {\cal K}}_{+}$ 
(resp. ${\hat {\cal K}_{-}}$) to be the subspace of  ${\hat {\cal K}}$ spanned by the 
eigenvectors of  ${\Sigma}_{z}$ with positive (resp. negative) eigenvalues. Thus, 
denoting by ${\hat {\Psi}}$ the simultaneous eigenvector of the ${\sigma}_{n,z}$\rq s 
with eigenvalues all equal to $-1$, ${\hat {\cal K}}_{\pm}$ are the  subspaces of ${\hat 
{\cal K}}$ generated by application to ${\hat {\Psi}}$ of the monomials of order greater 
than $L$ and less than $(L+1)$, respectively, in the different ${\sigma}_{n,x}$\rq s (or 
${\sigma}_{n,y}$\rq s). We denote their projection operators by ${\hat {\Pi}}_{\pm}$,  
respectively. We then define the phase cells ${\cal K}_{\pm}$ to be the subspaces 
${\tilde {\cal K}}{\otimes}{\hat {\cal K}}_{\pm}$ of ${\cal K}$, and denote their 
respective projection operators by ${\Pi}_{\pm} \ (=I_{\tilde {\cal K}}{\otimes}
{\hat {\Pi}}_{\pm})$. 
\vskip 0.2cm
Evidently, the formulation of the subspaces ${\cal K}_{\pm}$ of ${\cal K}$ here 
corresponds to that of  the previous Sections, with ${\alpha}$ taking the values $+$ and 
$-$, and fulfills the conditions of Eqs. (2.14) and (2.15). In the treatment that follows, we 
shall take the correspondence between the phase cells of  ${\cal I}$ and the eigenstates of 
$S$  to be the mapping  $r{\rightarrow}a(r)$ of  Section 3, with $a({\pm})={\pm}$. 
Thus, the phase cells ${\cal K}_{\pm}$ are the indicators for the vector states 
$u_{\pm}$, respectively. 
\vskip 0.3cm
{\bf 4.5. Ideality and Normality Conditions for ${\cal I}$.} It follows now the definition 
of ${\Pi}_{\pm}:=I_{\tilde {\cal K}}{\otimes}{\hat {\Pi}}_{\pm}$ that, on translating 
the ideality and normality conditions (3.2)$^{\prime}$ and (3.10)$^{\prime}$, 
respectively, into the specifications for this model and using Eqs. (2.23)-(2.26), the 
former condition reduces to the equation
$${\rm Tr}({\hat {\Omega}}_{+}{\hat {\Pi}}_{-})= {\rm Tr}({\hat {\Omega}}_{-}
{\hat {\Pi}}_{+})=0\eqno(4.22))$$
and the latter to
$$0<Max\bigl[{\rm Tr}({\hat {\Omega}}_{+}{\hat {\Pi}}_{-}), \ 
{\rm Tr}({\hat {\Omega}}_{-}{\hat {\Pi}}_{+})\bigr]<{\hat {\eta}}(L),\eqno(4.23)$$
where
$${\hat {\eta}}(L):={\eta}(2L+1).\eqno(4.24)$$
\vskip 0.3cm
{\bf 4.6. Resultant Properties of ${\cal I}$. } The following proposition establish that 
${\cal I}$ is an ideal measuring instrument for certain special values of the parameters of 
the model $S_{c}$ and is a normal one for a wide range of those parameters. Further, in 
the latter case, ${\hat {\eta}}(L)$ is exponentially small, i.e. of the order of 
${\rm exp}(-cL)$, with $c$ a positive constant of the order of unity. 
\vskip 0.3cm 
{\bf Proposition 4.1.} {\it Assuming  the conditions of Eq. (4.15), ${\cal I}$ has the 
following properties.
\vskip 0.2cm\noindent
(a)  If $J={\pi}/2$ and $m=1$, then ${\cal I}$  is an ideal instrument, with critical time 
${\tau}$. However, although this implies that it satisfies the local stability condition (I.3), 
it is transformed to a normal instrument by small perturbations of the global polarization 
$m$. 
\vskip 0.2cm\noindent
(b) If  $J{\in}({\pi}/4,{\pi}/2)$ and $m{\in}(-1,0)$, then ${\cal I}$ is a normal 
instrument, again with critical time ${\tau}$ and with ${\hat {\eta}}(L)={\rm exp}(-
cL)$, where $c$ is a numerical constant of the order of unity: specifically 
$c=-(1/2){\rm ln}(1-m^{2}{\rm cos}^{2}(2J))$. Moreover, in this case, the instrument is 
stable both under small perturbations of the global polarisation, $m$, and under local 
modifications of state.}
\vskip 0.3cm
It follows from our specifications that this proposition is a consequence of the following 
ones, which we shall prove below.
\vskip 0.3cm
{\bf  Proposition 4.2.} {\it  Assuming the conditions of Eq. (4.15), the model possesses 
the following properties.
\vskip 0.2cm\noindent
(i) If $J={\pi}/2$ and $m=1$, it satifisfies the ideality condition (4.22), with critical time 
${\tau}$.
\vskip 0.2cm\noindent
(ii) If $J{\in}({\pi}/4,{\pi}/2)$ and $m{\in}(0,1)$, it fulfills the normality condition 
(4.23), with critical time ${\tau}$ and
${\hat {\eta}}(L)=\bigl(1-m^{2}{\rm cos}^{2}(2J)\bigr)^{L/2}$.}   
\vskip 0.3cm
{\bf Proposition 4.3.} {\it  Assuming the conditions (4.15), 
\vskip 0.2cm\noindent
(i) the results of Prop. 4.1 are stable under any modification of the initial state 
${\hat {\Omega}}$ of ${\cal C}$ that is confined to some segment of this chain whose 
length is $O(1)$ with respect to the 'large' length $L$; 
\vskip 0.2cm\noindent
(ii)  under the conditions of Prop. 4.2 (i), any small perturbations of the global 
polarization $m$ change ${\cal I}$ from an ideal instrument to a normal one; and 
\vskip 0.2cm\noindent
(iii)  under the conditions of Prop. 4.2 (ii), the normality of the instrument is stable under 
small perturbations of the global polarization $m$.}
 \vskip 0.3cm
{\bf Proof of  of Proposition 4.2.} Let $v_{n,{\pm}}$ denote the eigenstate of 
${\sigma}_{n,z}$ whose eigenvalue is ${\pm}1$. Then, by definition of 
${\hat {\Pi}}_{+}$ (resp. ${\hat {\Pi}}_{-}$), the eigenstates of this projector are the 
tensor products of $n \ v_{-}$\rq s and $(2L+1-n) \ v_{+}$\rq s (resp. $n \ v_{+}$\rq s 
and $(2L+1-n) \ v_{-}$\rq s) with $n$ running  from $0$ to $L$. Hence, by Eqs. (4.20) 
and (4.21),
$${\rm Tr}({\hat {\Omega}}_{+}{\hat {\Pi}}_{-})=2^{-(2L+1)}{\sum}_{n=0}^{L} 
(1+m)^{n}(1-m)^{2L+1-n}(2L+1)!/n!(2L+1-n)!\eqno(4.25)$$
and
$${\rm Tr}({\hat {\Omega}}_{-}{\hat {\Pi}}_{+})=$$
$$2^{-(2L+1)}{\sum}_{n=0}^{L}
(1-(m){\rm cos}(2J))^{n}(1+(m){\rm cos}(2J))^{2L+1-n}
(2L+1)!/n!(2L+1-n)!
\eqno(4.26)$$
It follows immediately from these equations that, in the case where  $m=1$ and $J=
{\pi}/2$, the r.h.s.\rq s of these last two equations vanish. This completes the proof of 
Part (i) of the proposition.
\vskip 0.2cm
In order to prove Part (ii), we assume that $J{\in}({\pi}/4,{\pi}/2)$ and $0<m<1$. In this 
case, the summands on the r.h.s\rq s  of Eqs. (4.25) and (4.26) are positive for all 
$n{\in}[0,L]$, and they take their largest values at $n=L$, since 
$(2L+1)!/n!(2L+1-n)!, \  (1+m)^{n}(1-m)^{2L+1-m}$ and 
$(1-(m){\rm cos}(2J))^{n}(1+(m){\rm cos}(2J))^{2L+1-n}$ are all maximized at this 
value of $n$. Hence 
$$0<{\rm Tr}({\hat {\Omega}}_{+}{\hat {\Pi}}_{-}){\leq}
2^{-(2L+1)}(1+m)^{L}(1-m)^{L+1}(2L+1)!/(L!)^{2}\eqno(4.27)$$
and
$$0<{\rm Tr}({\hat {\Omega}}_{-}{\hat {\Pi}}_{+}){\leq}
2^{-(2L+1)}(1-(m){\rm cos}(2J))^{L}(1+(m)
{\rm cos}(2J))^{L+1}(2L+1)!/(L!)^{2}.(4.28)$$
Further since, by Sterling's formula,
$${\rm ln}\bigl[(2L+1)!/(L!)^{2}\bigr]={\rm ln}(L)+O(1),$$
it follows from Eqs. (4.27) and (4.28) that 
$${\rm ln}\bigl[{\rm Tr}({\hat {\Omega}}_{+}{\hat {\Pi}}_{-}\bigr]{\leq}
L{\rm ln}(1-m^{2})+{\rm ln}(L)+O(1)$$
and 
$${\rm ln}\bigl[{\rm Tr}({\hat {\Omega}}_{-}{\Pi}_{+}\bigr]{\leq}
L{\rm ln}(1-m^{2}{\rm cos}^{2}(2J))+{\rm ln}(L)+O(1)$$
Consequently, for sufficiently large $L$, the r.h.s.\rq s of these inequalities are both 
majorized by $(L/2){\rm ln}(1-m^{2}{\rm cos}^{2}(2J))$, and consequently, in view of 
the first parts of the inequalities (4.27) and (4.28), the normality condition (4.23) is 
fulfilled with ${\hat {\eta}}(L)={\rm exp}(-cL)$ and 
$c=-(1/2){\rm ln}\bigl(1-m^{2}{\rm cos}^{2}(2J)\bigr)$.
\vskip 0.3cm 
{\bf Proof of Proposition 4.3.} First consider the question of stability against global 
perturbations of the initial state corresponding to small changes in the polarisation $m$ 
which leave this parameter in the range $(0,1]$. In fact, it follows immediately from 
Prop. 4.2  that the normality condition (4.23) is stable under such perturbations, while  
the ideality condition (4.22) is changed to that of normality. This establishes Parts (ii) and 
(iii) of Prop. 4.3. 
\vskip 0.2cm
In order to prove Part (i), we introduce an arbitrary subset,  $K$, of  $[1,2L+1]$ whose 
total number of sites, ${\vert}K{\vert}$, is $O(1)$ with respect to the \lq large\rq\ length 
$L$; and we denote by  $K^{c}$ the complementary subset $[1,2L+1]{\backslash}K$. 
Correspondingly we denote by ${\check {\cal K}}$ and ${\check {\cal K}}^{c}$ the 
representation spaces for the spins in $K$ and $K^{c}$. It follows from this definition 
and that of ${\hat {\cal K}}$ that this latter space is the tensor product 
${\check {\cal K}}{\otimes}{\check {\cal K}}^{c}$.
\vskip 0.2cm
We now let ${\hat {\Omega}}_{1}$ be an arbitrary state of ${\cal C}$ that coincides 
with ${\hat {\Omega}}$ in $K^{c}$. Thus
$${\rm Tr}_{{\check {\cal K}}}({\hat {\Omega}}_{1})=
{\rm Tr}_{{\check {\cal K}}}({\hat {\Omega}}).\eqno(4.30)$$ 
By Eq. (4.19), the evolutes ${\hat {\Omega}}_{1,{\pm}}$ of ${\hat {\Omega}}_{1}$ 
that stem from the coupling of ${\cal I}$ to the states $u_{\pm}$ of $S$ are given by the 
formulae 
$${\hat {\Omega}}_{1,+}={\hat {\Omega}}_{1} \ {\rm and} \ 
{\hat {\Omega}}_{1,-}=Z^{\star}{\hat {\Omega}}_{1}Z.\eqno(4.31)$$
\vskip 0.2cm
We now need to show that the states ${\hat {\Omega}}_{1,{\pm}}$ satisfy the same 
condition (4.22) or (4.23) as ${\hat {\Omega}}_{\pm}$, and with the same value of 
${\hat {\eta}}(L)$, according to whether the assumptions of Prop. 4.2(i) or Prop. 4.2(ii) 
prevail. To this end, we note that, by Eqs. (4.3), (4.4), (4.30) and (4.31),
$$Tr_{{\check {\cal K}}}{\hat {\Omega}}_{1,{\pm}}=
Tr_{{\check {\cal K}}}{\hat {\Omega}}_{\pm}:=
{\check {\Omega}}_{\pm}^{c};\eqno(4.32)$$
and further, by Eqs. (4.16), (4.19) and (4.32), the states ${\check {\Omega}}_{\pm}$ are 
given by the following canonical analogues of Eqs. (4.20) and (4.21).
$${\check {\Omega}}_{+}^{c}=2^{-(2L+1)}
{\otimes}_{n{\in}K^{c}}(I_{n}+m{\sigma}_{n,z})\eqno(4.33)$$ 
and 
$${\check {\Omega}}_{-}^{c}=2^{-(2L+1)}{\otimes}_{n{\in}K^{c}}
\bigl(I_{n}+m{\sigma}_{n,z}{\rm cos}(2J)+m{\sigma}_{n,y}{\rm sin}(2J)\bigr).
\eqno(4.34)$$
 \vskip 0.2cm
We now denote by ${\check P}_{\pm}^{c}$ be the projection operators for the 
simultaneous eigenvectors of ${\lbrace}{\sigma}_{n,z}{\vert}n{\in}K^{c}{\rbrace}$ 
with eigenvalues all equal to ${\pm}1$, respectively; and by ${\check {\Pi}}_{+}^{c}$ 
(resp. ${\check {\Pi}}_{-}^{c}$) the projection operator of the subspace of  
${\check {\cal K}}^{c}$ for which ${\sum}_{n{\in}K^{c}}{\sigma}_{n,z}{\leq}-
{\vert}K{\vert} \ ({\rm resp.}{\geq}{\vert}K{\vert})$. It follows from these definitions 
that 
$$I_{{\check {\cal K}}}{\otimes}{\check P}_{\pm}^{c}<
{\hat {\Pi}}_{\pm}<
I_{{\check {\cal K}}}{\otimes}{\check {\Pi}}_{\pm}^{c}.\eqno(4.35)$$
Hence, by Eqs. (4.32) and (4.35), 
$${\rm Tr}_{{\check {\cal K}}^{c}}({\check {\Omega}}_{\pm}^{c}
{\check P}_{\mp}^{c}){\leq}
{\rm Tr}({\hat {\Omega}}_{1{\pm}}{\hat {\Pi}}_{\mp}){\leq}
{\rm Tr}_{{\check {\cal K}}^{c}}({\check {\Omega}}_{\pm}^{c}
{\check {\Pi}}_{\mp}^{c}).\eqno(4.36)$$
Further, by Eqs.(4.33) and (4.34) and the definitions of  ${\check P}_{\pm}^{c}$ and 
${\check {\Pi}}_{\pm}^{c}$, 
$${\rm Tr}_{{\check {\cal K}}^{c}}({\check {\Omega}}_{+}^{c}
{\check P}_{-}^{c})=2^{-(2L+1)}(1-m)^{2L+1-{\vert}K{\vert}},\eqno(4.37)$$
$${\rm Tr}_{{\check {\cal K}}^{c}}({\check {\Omega}}_{-}^{c}{\check P}_{+}^{c})
=2^{-(2L+1)}(1+(m){\rm cos}(2J))^{2L+1-{\vert}K{\vert}},\eqno(4.38)$$
$${\rm Tr}_{{\check {\cal K}}^{c}}({\check {\Omega}}_{+}^{c}
{\check {\Pi}}_{-}^{c})={\sum}_{n=0}^{L}(1+m)^{n}(1-m)^{2L+1-n-
{\vert}K{\vert}}
(2L+1-n-{\vert}K{\vert})!/n!(2L+1-n-{\vert}K{\vert})!\eqno(4.39)$$
and
$${\rm Tr}_{{\check {\cal K}}^{c}}({\check {\Omega}}_{-}^{c}
{\check {\Pi}}_{+}^{c})={\sum}_{n=0}^{L}$$
$$(1-(m){\rm cos}(2J))^{n}
(1+(m){\rm cos}(2J)))^{2L+1-n-{\vert}K{\vert}}
(2L+1-n-{\vert}K{\vert})!/n!(2L+1-n-{\vert}K{\vert})!.\eqno(4.40)$$
\vskip 0.2cm
Now in the ideal case where $m=1$ and $J={\pi}/2$, the r.h.s.\rq s Eqs. (4.37)-(4.40) all 
vanish. Therefore, in this case,  the two-sided inequalities (4.36) signify that 
${\rm Tr}({\hat {\Omega}}_{1,{\pm}}{\hat {\Pi}}_{\mp})$ vanishes, i.e. that the 
locally modified state ${\Omega}_{1}$ satifies the ideality conditions (4.22).
\vskip 0.2cm
In the normal case, where $m{\in}(0,1)$ and $J{\in}({\pi}/4,{\pi}/2)$, we see 
immediately from Eqs. (4.36)-(4.38) that the quantities 
${\rm Tr}({\hat {\Omega}}_{\pm}{\hat {\Pi}}_{\mp})$ are strictly positive. In order to 
obtain upper bounds for them, we treat Eqs. (4.39) and (4.40) by the method employed 
for the derivation of the estimates (4.28) and (4.29) from (4.25) and (4.26) in the proof of 
Prop. 4.2. Thus, taking account of the demand that ${\vert}K{\vert}=O(1)$ with respect 
to $L$, we obtain precisely the same estimates for 
${\rm Tr}({\hat {\Omega}}_{1,{\pm}}{\hat {\Pi}}_{\mp})$ as  those given by Eqs. 
(4.28) and (4.29) for  ${\rm Tr}({\hat {\Omega}}_{\pm}{\hat {\Pi}}_{\mp})$. This 
signifies that ${\cal I}$ remains a normal instrument, with unchanged value of ${\hat 
{\eta}}(L)$, when the initial state of ${\cal C}$ is changed from ${\hat {\Omega}}$ to 
${\hat {\Omega}}_{1}$. In other words, the operation of the instrument ${\cal I}$ is 
stable under local modifications of the initial state of  the chain ${\cal C}$.
\vskip 0.5cm
\centerline {\bf  5. Concluding Remarks}
\vskip 0.3cm
We have shown that the general scheme of Sections 2 and 3 is fully realized by the model 
of Section 4. Since that is a Hamiltonian model for the composite $S_{c}$ of 
microsystem and measuring instrument, this signifies that the traditional quantum 
mechanics of finite conservative systems provides a perfectly adequate framework for the 
quantum theory of measurement. This  theory therefore requires no extraneous elements, 
such as the interaction of  $S_{c}$ with the \lq rest of the Universe\rq\  or a nonlinear 
modification of  its Schroedinger dynamics, as has been proposed by some authors [4-7]. 
Furthermore, the treatment of the model of Section  provides a clear illustration of the 
mathematical dichotomy of ideal and normal measuring instruments. It also establishes 
that, from an empirical standpoint, there is effectively no distinction between these two 
classes of instruments, since the odds against the indication by a normal instrument of a 
\lq wrong\rq\ state of  a the microsystem are truly astronomical, being of the order of 
${\rm exp}(cL)$ to one, where $c$ is of the order of unity. 
\vskip 0.2cm
This last observation implies that the instrument of the Coleman-Hepp model would work 
perfectly well if it the chain ${\cal C}$ were merely mesoscopic rather than macroscopic, 
e.g. with the chain ${\cal C}$ composed of, say, $10^{5}$ spins. This raises the 
question, that can be addressed both experimentally and by the study of other models, of 
whether real quantum measuring instruments of mesoscopic size can be devised.
 \vskip 0.5cm
{\bf Acknowledgments.} It is a pleasure to thank Professors Gerard Emch and Nico Van 
Kampen for their constructive comments on an earlier draft of this article.
\vskip 0.5cm
\centerline {\bf References.} 
\vskip 0.3cm\noindent
[1] J. Von Neumann: {\it Mathematical Foundations of Quantum Mechanics}, Princeton 
University Press, Princeton, NJ, 195
\vskip 0.2cm\noindent
[2] N. G. Van Kampen: {\it Physica} A {\bf 153}, 97 (1988).
\vskip 0.2cm\noindent
[3]  N. Gisin: {\it Phys. Rev. Lett.} {\bf 52}, 1657 (1984).
\vskip 0.2cm\noindent
[4] E. Joos and H. D. Zeh: {\it Z. Phys.} B {\bf 59}, 223 (1985).
\vskip 0.2cm\noindent
[5] L. Diosi: {\it J. Phys.} A {\bf 21}, 2885 (1988).
\vskip 0.2cm\noindent
[6] I. Percival: {\it Quantum State Diffusion}, Cambridge Univ. Press,  Cambridge, 1998.
\vskip 0.2cm\noindent
[7] G. C. Ghirardi, A. Rimini and T. Weber: {\it Phys. Rev.} D {\bf 34}, 470 (1986).
\vskip 0.2cm\noindent 
[8] N. Bohr: {\it Discussion with Einstein on epistomological problems in atomic 
physics}, Pp. 200-241 of  {\it Albert Einstein: Philosopher-Scientist}, Ed. P. A. Schilp, 
The Library of Living Philosophers, Evanston, IL, 1949.
\vskip 0.2cm\noindent
[9] J, M. Jauch: {\it Foundations of Quantum Mechanics}, Addison Wesley, Reading, 
MA, 1968.
\vskip 0.2cm\noindent
[10] B. Whitten-Wolfe and G. G. Emch: {\it Helv. Phys. Acta} {\bf 49}, 45 (1976).
\vskip 0.2cm\noindent
[11] G. G. Emch: Pp. 255-264 of  {\it Quantum Information and Communication},  E. 
Donkor, A. R. Pirich and H. E. Brandt, Eds., Intern. Soc. Opt. Eng. (SPIE) Proceedings 
5105 (2003).
\vskip 0.2cm\noindent
[12] N. G. Van Kampen: {\it Physica} {\bf 20}, 603 (1954).
\vskip 0.2cm\noindent
[13] G. G. Emch: {\it Helv. Phys. Acta} {\bf 37}, 532 (1964).
\vskip 0.2cm\noindent
[14] K.Hepp: {\it Helv. Phys. Acta} {\bf 45}, 237 (1972).
\vskip 0.2cm\noindent
[15] D. Ruelle: {\it Statistical Mechanics}, W. A. Benjamin, New York, 1969.
\vskip 0.2cm\noindent
[16] G. G. Emch: {\it Algebraic methods in Statistical Mechanics and Quantum Field 
Theory},  Wiley, New York, 1972.
\vskip 0.2cm\noindent
[17] G. L. Sewell: {\it Quantum Mechanics and its Emergent Macrophysics}, Princeton 
University Press, Princeton, 2002.
\vskip 0.2cm\noindent
[18]  J. S. Bell: {\it Helv. Phys. Acta} {\bf 48}, 93 (1975).
\vskip 0.2cm\noindent
[19] E. P.Wigner: {\it Symmetries and Reflections}, Indiana University Press, 
Bloomington, 1967.
\end